\documentclass[10pt,a4paper]{article}

\usepackage{times}

\usepackage{epsfig,wrapfig}
\usepackage{epstopdf} 

\usepackage{fancyhdr}

\usepackage[lmargin=2.5cm, rmargin=2.5cm,tmargin=3.50cm,bmargin=3.50cm]{geometry}
\usepackage{indentfirst}
\usepackage{multicol}
\usepackage{blindtext}
\usepackage{titlesec}
\titleformat{\section}[hang]
  {\centering}{\thesection}{1ex}{\normalsize \textsc}
\titleformat{\subsection}[hang]
  {}{\thesubsection}{1ex}{\normalsize \textit}

\newcommand{\acknowledgement}{\section*{\centering{\textnormal{\normalsize{\textsc{Acknowledgement}}}}}}

\renewcommand{\thesection}{ \normalsize \textnormal{\Roman{section}.}}
\renewcommand{\thesubsection}{\normalsize \textnormal{\textsc{\textit{\Alph{subsection}.}}}}

\def\e{\begin{equation}}
\def\f{\end{equation}}
\def\_#1{{\bf #1}}

\def\.{\cdot}

\begin{document}

\title{\large \textbf{Broadband multilayer metasurface absorbers with MXene resonators and topology optimized substrates \footnote{Copyright (C) 2025 IEEE. Personal use of this material is permitted.  Permission from IEEE must be obtained for all other uses, in any current or future media, including reprinting/republishing this material for advertising or promotional purposes, creating new collective works, for resale or redistribution to servers or lists, or reuse of any copyrighted component of this work in other works.}}}
  
%
\def\affil#1{\begin{itemize} \item[] #1 \end{itemize}}
\author{\normalsize \bfseries \underline{M.T. Passia}$^{1}$, Y.Zhao$^1$, H.Wang$^1$  and S.A.Cummer$^1$
}
\date{}
\maketitle
\thispagestyle{fancy} 
\vspace{-6ex}
\affil{\begin{center}\normalsize $^1$Duke University, Department of Electrical and Computer Engineering, 27708, Durham, NC, USA \\
mariathaleia.passia@duke.edu
 \end{center}}

\begin{abstract}
\noindent \normalsize
\textbf{\textit{Abstract} \ \ -- \ \
We present the synthesis of broadband multilayer metamaterial absorbers (MMA) based on MXenes, which are novel two-dimensional conductive materials with higher ohmic losses than copper.  MXene resonator of different conductivity can be implemented at each layer or across the same layer, offering increased design flexibility.  We examine the possibility of utilizing topology-optimized stereolithography (SLA) 3D-printed substrates as a complementary means for enhancing absorption. Combining MXenes with topology optimized 3D-printable structures paves the way for realizing MMAs of enhanced absorption.
}
\end{abstract}

\section{Introduction}

Broadband absorption is essential in many microwave applications, such as RCS reduction and electromagnetic interference shielding. To enhance the absorption bandwidth, non-uniform multi-element MMAs have been introduced. Individual elements of the MMA resonate at different frequencies, leading to an extended absorption bandwidth. 
The resonators of a non-uniform MMA can be located on the same plane or on different planes, forming a multilayer absorber~[1]. 
Advancements in 3D printing technology have enabled the realization of various easily manufacturable devices. Filament-based 3D printing has been mostly used for the fabrication of MMAs~[2]. Multilayer 3D-printed MMAs have been realized, with conductive and dielectric filaments interleaved to form the multilayer stack~[3]. However, existing multilayer MMAs have substrates with a constant or pyramid-shaped dielectric material. 
Recently, MXenes have been utilized to design MMAs~[4,5]. MXenes are a novel 2D conductive material that can be used instead of metal-based inks or filaments. MXenes have considerably lower conductivity than copper and, in turn, higher ohmic losses, which are beneficial for MMAs. MXenes can be synthesized to have a variety of conductivity values.
A plasma-assisted atomic layer etching (ALE) process~[6] can modify the MXene's surface chemistry and increase the MXene's conductivity.
Adjusting the MXene conductivity offers additional design flexibility  compared to other metal alternatives. 
In this work, we synthesize multilayer MXene MMAs and examine the potential of using topology-optimized substrates that can be fabricated by high-precision SLA 3D-printing.  Topology-optimized substrates can be used as a complementary means to tailor the metal resonators. Combining MXenes with topology-optimized 3D-printable substrates paves the way for enhancing MMA absorption.

\section{MXene-based multilayer MMA}
We synthesize a PEC-backed MXene-based MMA consisting of six dielectric and six metal interleaved layers, as shown in Fig.1(a). Each dielectric layer has a thickness of $h$~=~0.6~mm and an initial dielectric constant of $\epsilon_r~=~1.25(1-j0.03)$. Each metal layer consists of square MXene patches of a thickness in the range of a few micrometers. The MXene patch length $W_i$ varies across different layers, with $W_i$~=~[10.1, 9.14, 8.18, 7.22, 6.26, 5.3]~mm,  starting from the lower layer. By considering patches of a different size across each layer, we position resonances to span the desirable frequency range.   We have conducted preliminary measurements of uniform 3~um MXene samples deposited on a 20~um polycarbonate membrane in the X-band. The MXene sample is placed between two WR90 sections to perform the S-parameter measurements using a VNA~[7]. The bulk AC conductivity is estimated close to $\sigma_M~=~5\times10^4$~S/m for the non-plasma-treated sample. Higher conductivity values up to about $\sigma_M~=~11\times10^4$~S/m are achieved for plasma-treated MXenes~[6].
MXenes have a considerably lower conductivity than copper ($\sigma_C~=~5.88\times10^7$~S/m), making MXenes a favorable alternative for synthesizing MMAs, as they offer increased ohmic losses. We simulate the MMA of Fig.1(a) in Comsol Multiphysics with non-plasma-treated MXenes selected across all layers. As we aim to measure the device in a rectangular waveguide,  the transverse dimensions of the device $L_x$, $L_y$ are the inner dimensions of a WR90 waveguide. We apply PEC boundary conditions on side boundaries and excite the device by a TE$_{10}$ mode. The absorbance is shown in Fig.1(b) between 10 GHz and 19 GHz. The multilayer MXene-based MMA has an average absorbance of 91\%.

Absorption can be further enhanced  by  optimizing the MXene  or the dielectric layers. Regarding the MXenes, we can change their shape, for example, by topology optimization techniques or by considering different conductivity values across patches.  By using high-precision SLA 3D-printing to fabricate the MMA substrates, we can also  perform a topology optimization on the dielectric substrates, where each cell  can be either air or resin. Each substrate cell is set to approximately 0.5~mm in the transverse plane, adhering to the 3D-printer's specifications. The synthesis methodology comprises two interleaved optimization steps, one for the MXenes and the other for the substrates. In the next section, we will demonstrate the optimization step that involves the dielectric substrates.

  \begin{figure}
    \centering
    \hspace{-0.1cm}
    \begin{tabular}[b]{c} 
      \includegraphics[width = 0.4\columnwidth,  trim = {4cm 7.5cm 4cm 5cm}, clip]{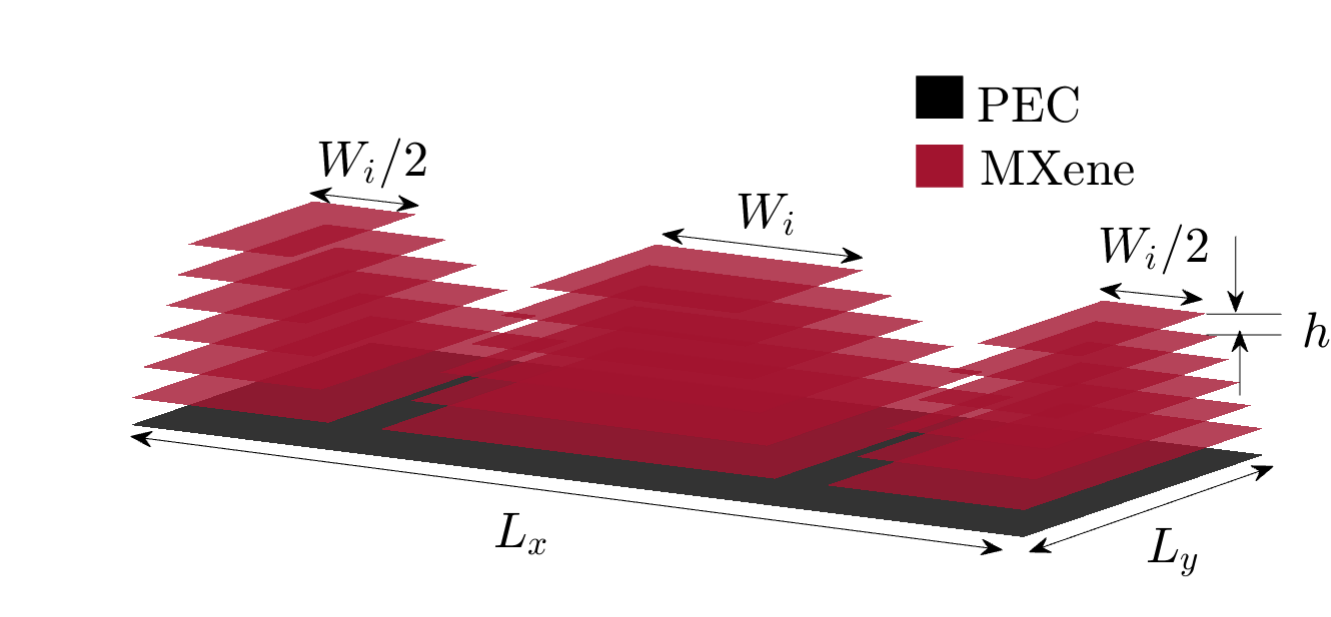}\\
      \small (a)
      \end{tabular}\hspace{-0.5cm}
      \begin{tabular}[b]{c}
        \includegraphics[width = 0.55\columnwidth,  trim = {0cm 0cm 0cm 0cm}, clip]{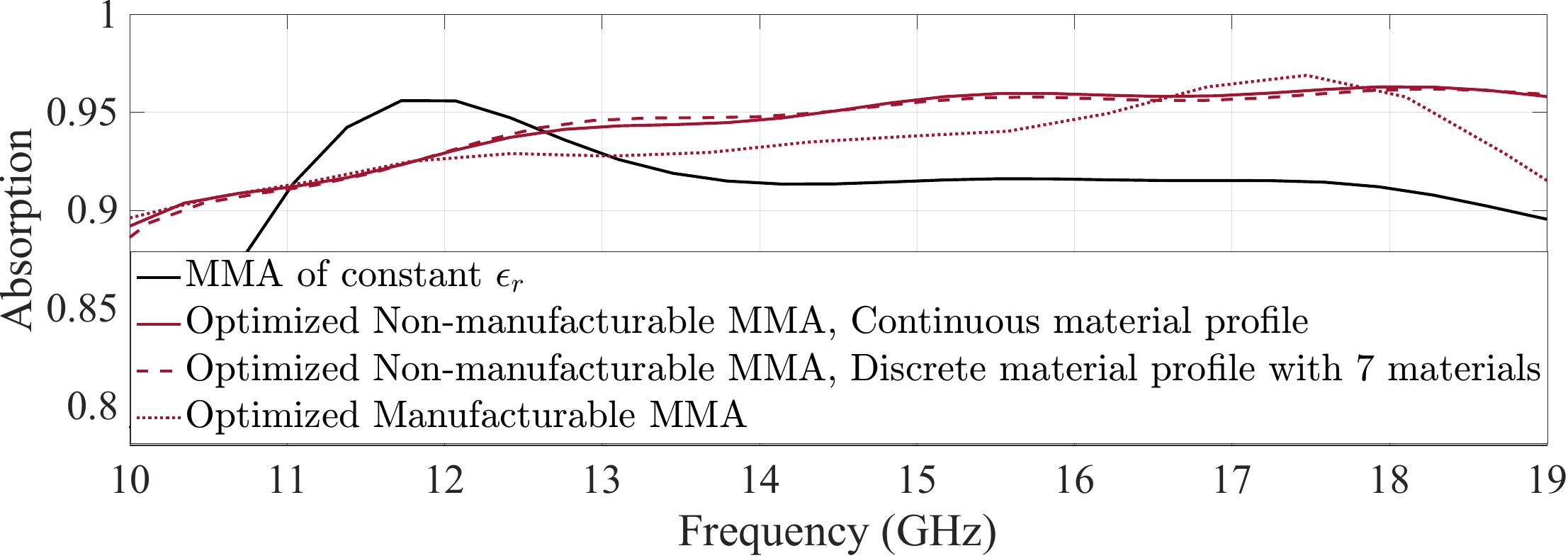}\\
      \small (b)
    \end{tabular}
    \caption{(a) The multilayer MMA with MXene patches arranged on dielectric substrates of thickness $h$.  (b) The MMA absorbance vs. frequency for (i) the  MMA with constant $\epsilon_r$,  the optimized non-manufacturable MMA with (ii) a continuous material profile (iii) 7 materials and (iv) the manufacturable MMA.}
    \label{fig:geom}
  \end{figure}
  

\section{Topology-optimized MMA}

We obtain an optimized inversely designed MMA by performing a density-based topology optimization with the TopOpt method~[8], implemented in Comsol Multiphysics. The TopOpt method uses adjoint sensitivity analysis to perform gradient computations, which requires only solving a single (adjoint) equation for the optimization objective and one for each constraint in the optimization problem, independent of the size of the design space. We aim at maximizing absorption over the desired frequency range.  Although fabrication constraints, such as having a binary dielectric constant profile of air and resin ($\epsilon_r=2.7(1-j0.03)$) can be added to the optimization, we choose to solve for a continuous dielectric constant profile. We  transform the inversely designed continuous dielectric profile to a binary and connected one that can be manufactured by SLA 3D printing, by employing the \texttt{BINACONN} methodology~[9].   We aim to preserve the MMA absorption as close as possible to the continuous case. The \texttt{BINACONN} methodology  includes the following steps: (i) the continuous profile is transformed into a discrete one by preserving as many material levels needed, so that the device's performance remains reasonably close to the continuous case. (ii) We assign a resin percentage to each non-manufacturable material level instead of assigning a fixed air/dielectric structure. This provides additional flexibility, which is particularly useful for complex and conformal-shaped structures. (iii) We generate manufacturable connected devices of the prescribed resin percentage and substitute the non-manufacturable material levels to approximate the initial device. We generate multiple manufacturable devices and select the one with a performance as close as possible to the continuous case. 

 The MMA absorbance  of the MMA with $\epsilon_r=1.25(1-j0.03)$, the optimized non-manufacturable and manufacturable MMAs is shown in Fig.1(b) across the desired frequency range.  We observe that the non-manufacturable MMA of seven materials is almost identical to that of a continuous material profile.
 Both the optimized non-manufacturable and manufacturable MMAs have a higher absorption than the device of a constant material profile, by about 2.5\%-3.5\% on average.  
When implementing \texttt{BINACONN} we generated ten manufacturable devices and selected the one with the closest performance to the non-manufacturable device. Although additional manufacturable devices may be generated and tested to achieve an even higher level of agreement,  the selected binary device is satisfactorily close to the non-manufacturable optimized device for this initial synthesis stage.
 Fig.2(a) shows the material profiles of the non-manufacturable MMA with seven materials, whereas Fig.2(b) shows the profiles of the manufacturable MMA in the same order. We observe that the material profiles vary in the transverse plane. 
Each substrate is substituted by two material profiles of 0.3~mm thickness, in accordance with the 3D printer's specifications.  The total MMA will be assembled by extruding parts of each lower layer, with complementary gaps on the immediate upper layer, without the use of glue. The ultrathin MXene patches will also be inserted at appropriate positions, with the patch position indicated on the 3D-printed pattern.

\begin{figure}
  \centering
  \hspace{-0.1cm}
  \begin{tabular}[b]{c} 
    \includegraphics[width = 0.47\columnwidth,  trim = {0cm 1.5cm 0cm 1cm}, clip]{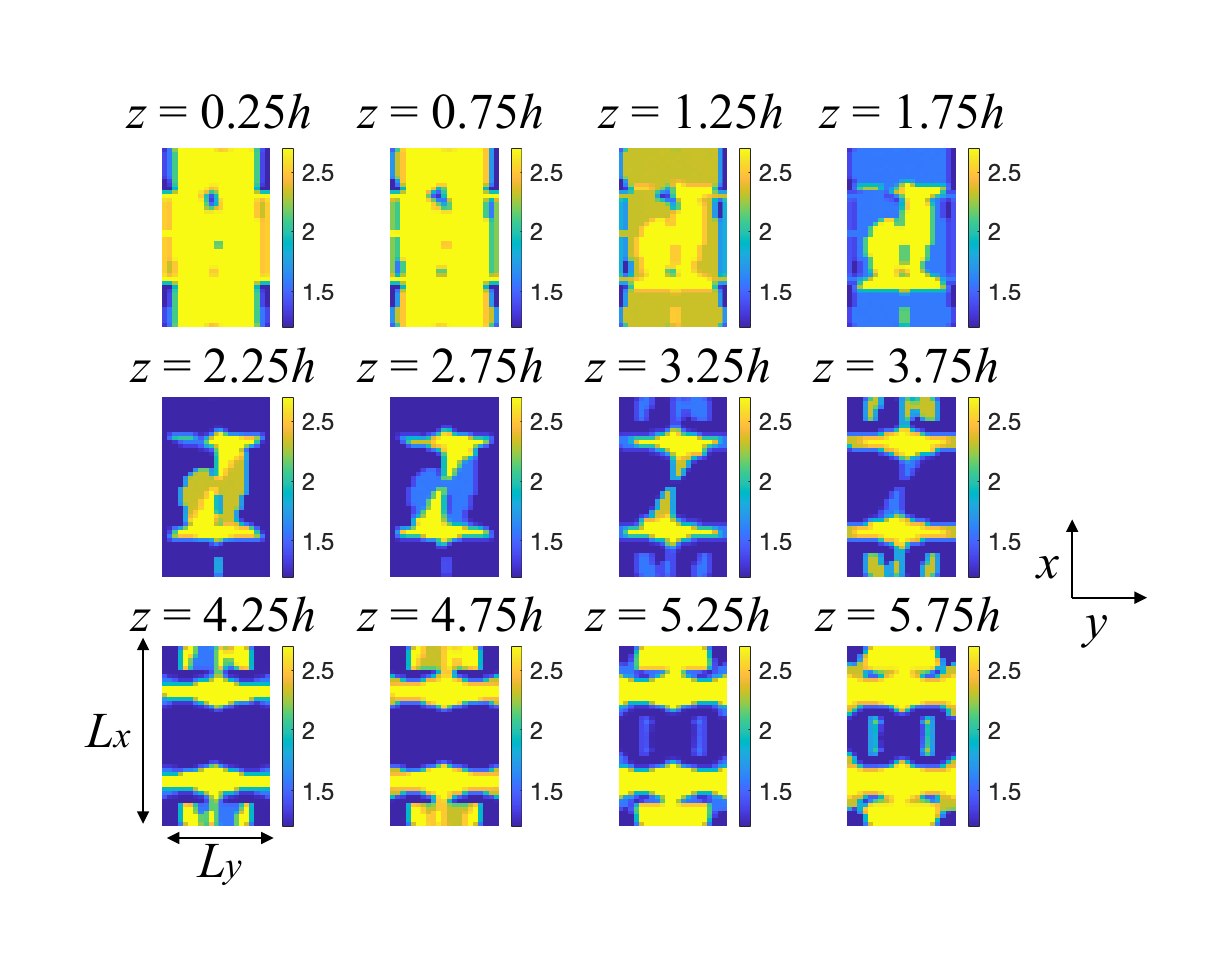}\\
    \small (a)
    \end{tabular}\hspace{-0.5cm}
    \begin{tabular}[b]{c}
      \includegraphics[width = 0.47\columnwidth,  trim = {0cm 1.1cm 0cm 0.5cm}, clip]{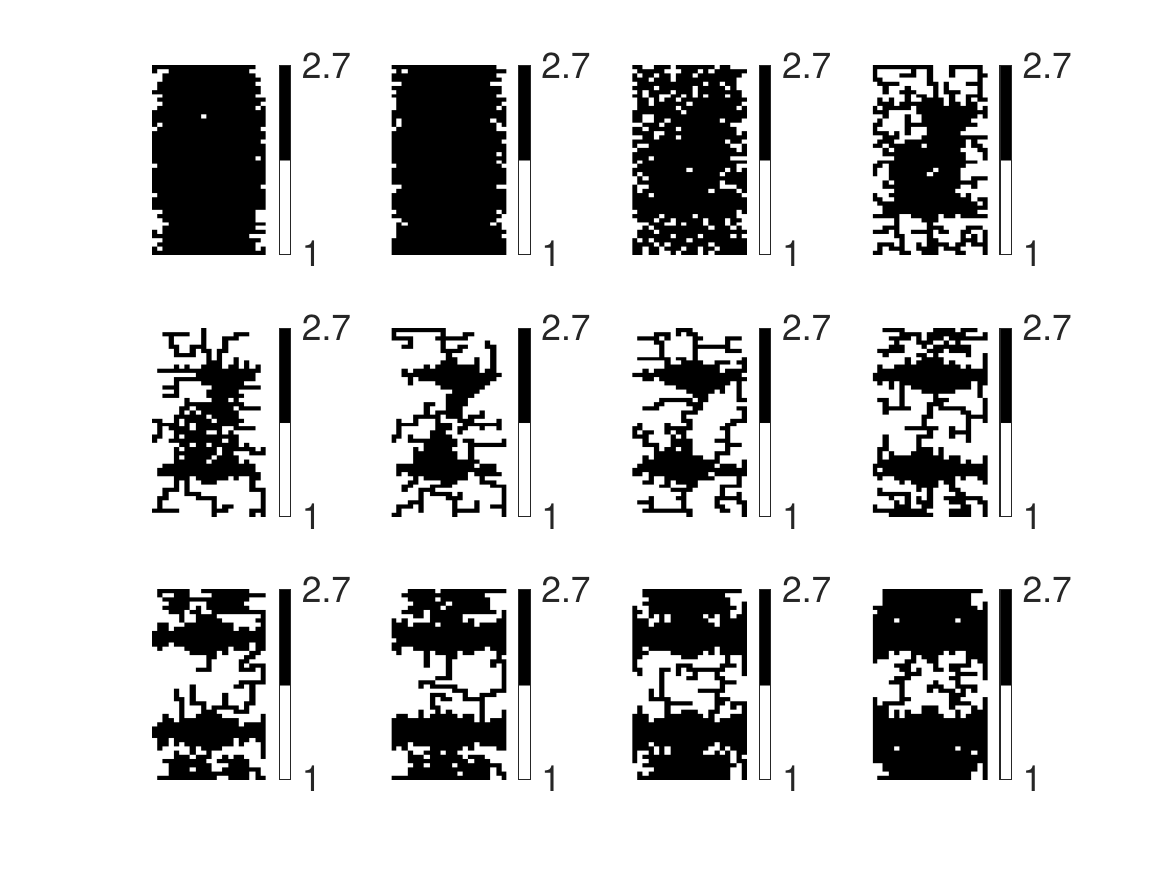}\\
    \small (b)
  \end{tabular}
  \caption{(a) Non-manufacturable 7-material and (b) Manufacturable material profiles.} 
  \label{fig:prof}
\end{figure}

\section{Conclusion}
We examined the synthesis of multilayer MMAs based on MXenes of increased ohmic losses, and inversely-designed 3D-printable substrates.   The latter further enhance the MMA's absorption and offer a complementary means to optimizing the metal structures for improving absorption.

\acknowledgement

This project has received funding from the European Union's Horizon 2020
research and innovation programme under the Marie Skłodowska-Curie grant
agreement No.101146306. 
\includegraphics[width = 0.25\columnwidth]{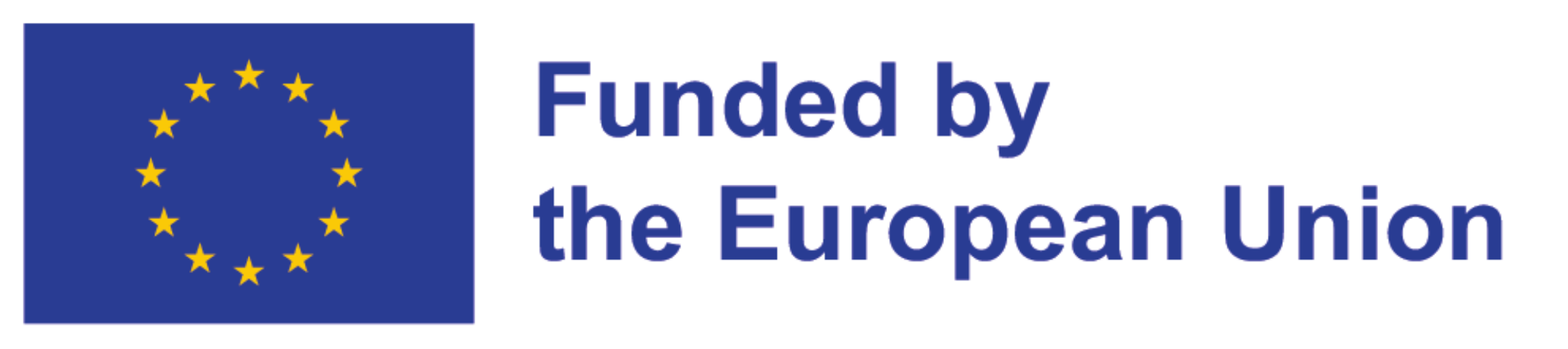}

\end{document}